\documentstyle{l-aa}     
\input psfig   
\newcommand{\asec}  {\hbox{$^{\prime\prime}$}}
\newcommand{\amin}  {\hbox{$^{\prime}$}}
\newcommand{\hI}    {H\,{\sc i} }
\newcommand{\hII}    {H\,{\sc ii} }
\newcommand{\kms}   {\hbox{${\rm km\,s}^{-1}$}}
\newcommand{\kmsmpc}{\hbox{${\rm km\,s}^{-1}{\rm Mpc}^{-1}$}}
\begin{document}
\topmargin=+2.0cm
\thesaurus{ 
	    09.11.1;  
            11.09.1;  
            11.09.1;  
            11.09.4;  
            11.11.1;  
            11.19.2;  
	         }
   \title{Vertical equilibrium of molecular gas in galaxies}
   \subtitle{}
   \author{F.~Combes and J-F.~Becquaert}
   \offprints{F.~Combes}
   \institute{ Observatoire de Paris, DEMIRM, 61 av. de l'Observatoire, 
F-75014 Paris}  
   \date{Received date; Accepted date}
   \maketitle
%
   \begin{abstract}
 We present CO(1-0) and CO(2-1) observations of the two nearly face-on
galaxies NGC 628 and NGC 3938, in particular cuts along the major and 
minor axis. The contribution of the beam-smeared in-plane velocity gradients
to the observed velocity width is quite small in the outer 
parts of the galaxies. This allows us to derive the velocity dispersion of 
the molecular gas perpendicular to the plane. We find that this
dispersion is remarkably constant with radius, 
6 \kms for NGC 628 and 8.5 \kms for NGC 3938, and of the same order
as the \hI\ dispersion. The constancy of the value is interpreted in terms of 
a feedback mechanism involving gravitational instabilities and
gas dissipation. The similarity of the CO and \hI\ dispersions suggests
that the two components are well mixed, and are only two different phases
of the same kinematical gas component. The gas can be transformed from the 
atomic phase to the molecular phase and vice-versa several 
times during a z-oscillation.
   \end{abstract}
\section{Introduction}
 An important diagnostic of the physical state of the interstellar medium
is its large-scale velocity dispersion. This parameter is however very 
difficult to derive, since it is in general dominated by the
contribution of the systematic velocity gradients in the beam, which are
not well-known.
Exactly face-on galaxies are ideal objects for this study, since
the line-width can be attributed almost entirely to the z-velocity
dispersion $\sigma_v$. Indeed, the systematic gradients perpendicular
to the plane are expected negligible; for instance no systematic pattern
associated to spiral arms have been observed in face-on galaxies
(e.g. Shostak \& van der Kruit 1984,  Dickey et al 1990), implying that
the z-streaming motions at the arm crossing are not predominant. In an 
inclined galaxy on the contrary, it is very difficult 
to obtain the true velocity
dispersion, since the systematic motions in the plane $z=0$ (rotation, arm
streaming motions) widen the spectra due to the finite spatial resolution of
the observations (e.g. Garcia-Burillo et al 1993, Vogel et al 1994). 

 Nearly face-on galaxies have already been extensively studied in the
atomic gas component, in order to derive the true \hI\ velocity dispersion
(van der Kruit \& Shostak
1982, 1984, Shostak \& van der Kruit 1984, Dickey et al 1990). 
The evolution of $\sigma_v$ as a function of radius was derived: the velocity
dispersion is remarkably constant all over the galaxy
$\sigma_v$ = 6 \kms = $\Delta V_{FWHM}/2.35$, and only in the
inner parts it increases up to 12 \kms.
\medskip

The constancy of $\sigma_v$ in the plane, and in particular in the outer 
parts of the galaxy disk, is not yet well understood; it might
be related to the large-scale gas stability and to the linear 
flaring of the plane,
as is observed in the Milky-Way (Merrifield 1992) and M31 (Brinks \& Burton
1984).  In the isothermal sheet model of a thin plane, where the z-velocity
dispersion $\sigma$ is independent of z, the height $h_g$(r) of the
gaseous plane, if assumed self-gravitating, is
$$
h_g(r) = {{\sigma_g^2(r)}\over{2 \pi G \mu_g(r)}}
$$
where $\sigma_g$ is the gas velocity dispersion, 
and $\mu_g(r)$ the gas surface
density. The density profile is then a sech$^2$ law.
 But to have the gas self-gravitating, we have to assume that either there is
no dark matter component, or the gas is the dark matter itself
(e.g. Pfenniger et al 1994). Since in general
the \hI\ surface density decreases as 1/r in the outer parts of galaxies
(e.g. Bosma 1981), a linear flaring ($h_g \propto r$) corresponds
to a constant velocity dispersion with radius.

On the contrary hypothesis
of the gas plane embedded in an external potential of larger scale
height, where the gravitational acceleration close to the plane can
be approximated by $K_z z$, the z-density profile is then a Gaussian:
$$
\rho_g=\rho_{0g} e^{-\frac12\frac{K_zz^2}{\sigma_g^2}}
$$
and the characteristic height, or gaussian scale height of the gas is:
$$
 h_g  =  \frac{\sigma_g}{\sqrt{K_z}}
$$
and $K_z$ is $4\pi G \rho_0$, where $\rho_0(r)$ is the density in the plane
of the total matter, stellar component plus dark matter component, in which the
gas is embedded. If the dark component is assumed spherical, the
density in the plane is dominated by the stellar component, which is
distributed in an exponential disk.
This hypothesis would predict an exponential flare in the gas,
while the gas flares appear more linear than exponential
(e.g. Merrifield 1992, Brinks \& Burton 1983). The knowledge of
their true shape is however hampered by the presence of warps.
Also, the flattening of the dark matter component, and its participation
to the density $\rho_0$ in the plane, is unknown.
\medskip

As for the stability arguments, let us assume
here the z-velocity dispersion
comparable to the radial velocity dispersion, or at least their
ratio constant with radius. 
The velocity dispersion of the gas component is self-regulated by
dynamical instabilities.  If the Toomre Q parameter for the gas
$$
Q_g = \sigma_g(r)  \kappa(r) /3.36 G \mu_g(r) = \sigma_g(r)/\sigma_{cg}(r)
$$
is lower than 1, instabilities set in, heat the medium and increase 
$\sigma_g(r)$ until $Q_g$ is 1. 
The critical velocity dispersion $\sigma_{cg}$ depends on the
epicyclic frequency $\kappa(r)$ and on the gas surface density $\mu_g(r)$; 
assuming again an \hI\ surface density decreasing as 1/r in the outer parts
and a flat rotation curve, where $\kappa(r)$ also varies as
1/r, then $\sigma_{cg}$ is constant. To maintain $Q_g = 1$ all over the outer
parts, $\sigma_g$ should also remain constant.

However, the gas density gradient appears often steeper than $1/r$ and
the $Q_g$ parameter is increasing towards the outer parts. This has
been noticed by Kennicutt (1989), who concluded that there exists some radius 
in every galaxy where the gas density reaches the threshold of global
instability ($Q_g\approx 1$); he identifies this radius to the onset
of star formation in the disk. In fact, this threshold does not occur
exactly at $Q_g$ = 1, but at a slightly higher value, around 1.4, which 
could be due to the fact that the $Q$ criterion is a single-fluid
one, which does not take into account the coupling between gas and stars. 

\medskip

The determination of the z-velocity dispersion 
in the molecular component has not yet been done. It could bring 
complementary insight to the \hI\ results,
since in general the center of galaxies is much better sampled 
through CO emission (a central \hI\ depletion
is frequent), and also the thickness of the H$_2$ plane can be lower by a
factor 3 or 4 than the \hI\ layer (case of MW, M31, Boulanger et al 1981). In
the case of M51, an almost face-on galaxy (i=20$^{\circ}$), the estimated
$\sigma_v$ determined from the CO lines is surprisingly large (up to
$\sigma_v$ = 25 \kms in the southern arm) once the rotation field, and even 
streaming-motions are taken into account, at the beam scale. 
An interpretation could be that the CO lines are broadened by
macroscopic opacity, i.e. cloud overlapping  (Garcia-Burillo et al 1993),
since such large line-widths are not observed in galaxies with less CO
emission. However, one could also suspect turbulent motions, generated at
large-scale by gravitational instabilities or viscous shear.
The level of star formation could be another factor: as for turbulence, it
generally affects the molecular component more than the HI, except for very
violent events like SNe. But the finite
inclination (20$^{\circ}$) of M51 makes the discrimination between in-plane
and z-dispersion very delicate. It is therefore necessary
 to investigate in more details this
problem in exactly face-on galaxies, and determine whether there exist spatial
variations of $\sigma_v$ over the galaxy plane.
\medskip

\begin{figure}
\psfig{figure=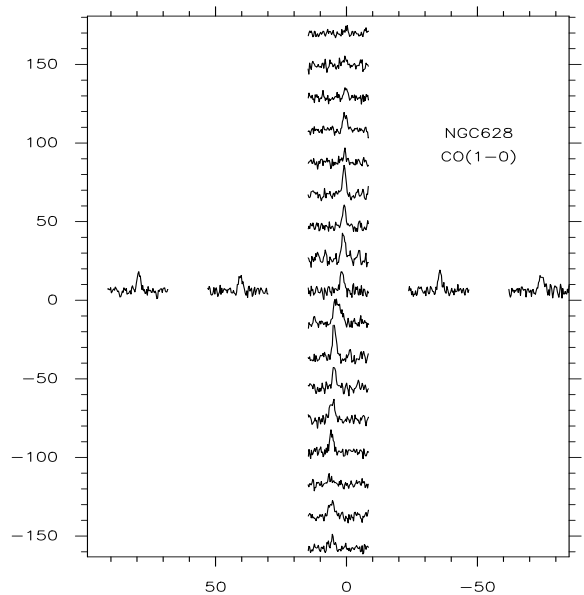,height=9.cm,width=9.cm}
\caption[]{Map of CO(1-0) spectra towards NGC 628. The scale in
velocity is from 556 to 756 \kms, and in T$_A^*$ from -0.05 to 0.18K.
 The major axis has been rotated by 25$^\circ$ to be vertical. The
offsets marked on the box are in arcsec.}
\label{map628}
\end{figure}

\begin{figure}
\psfig{figure=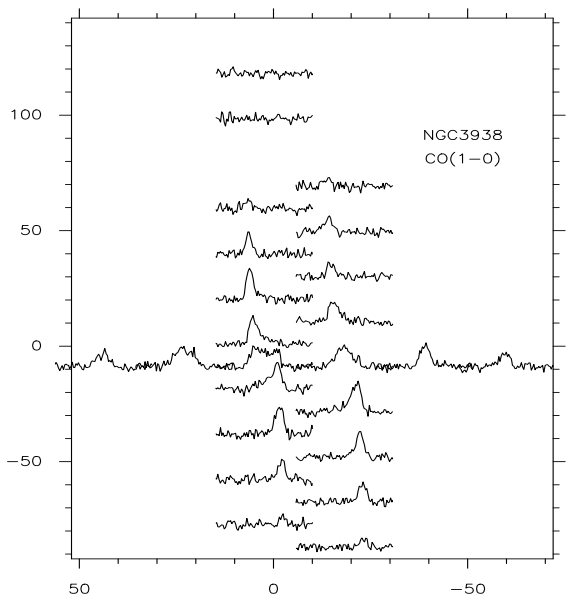,height=9.cm,width=9.cm}
\caption[]{Map of CO(1-0) spectra towards NGC 3938. The scale in
velocity is from 608 to 908 \kms, and in T$_A^*$ from -0.05 to 0.18K.
 The major axis has been rotated by 20$^\circ$ to be vertical. The
offsets marked on the box are in arcsec.}
\label{map3938}
\end{figure}

In this paper we report molecular gas observations of two face-on galaxies
NGC 628 (M74) and NGC 3938, in the CO(1-0), CO(2-1) and $^{13}$CO lines, 
using the IRAM 30--m telescope. After a brief description of the
galaxy parameters in section 2, and the observational parameters 
in section 3, we derive the amplitude and the spatial
variations of $\sigma_v$ perpendicular to the plane in NGC 628 and NGC 3938.
 Section 5 summarises  and discusses the physical interpretations.

\begin{table*}
\begin{flushleft}
\caption[]{Galaxy properties}
\scriptsize
\begin{tabular}{|lllrcccrrcccc|}
\hline
 & & & & & &  & & & & & & \\
\multicolumn{1}{|c}{Name}             &
\multicolumn{1}{c}{Type}              &
\multicolumn{2}{c}{Coordinates}       &
\multicolumn{1}{c}{$V_{\odot}$}       &
\multicolumn{1}{r}{Dist}          &
\multicolumn{1}{c}{$L_B$} 	      &
\multicolumn{1}{c}{f$_{60}$}     &
\multicolumn{1}{c}{f$_{100}$}    &
\multicolumn{1}{c}{$D_{25}$}          &
\multicolumn{1}{c}{$PA$}          &
\multicolumn{1}{c}{$i$}          &
\multicolumn{1}{c|}{Environment}      \\
 &   & \multicolumn{1}{c}{$\alpha$(1950)} &
\multicolumn{1}{c}{$\delta$(1950)}          &
\multicolumn{1}{c}{$km\,s^{-1}$}      &
\multicolumn{1}{c}{Mpc}               &
\multicolumn{1}{c}{10$^9$ $L_\odot$} 	      &
\multicolumn{2}{c}{Jy}                &
\multicolumn{1}{c}{$'$}               &
\multicolumn{1}{c}{$\circ$}               &
\multicolumn{1}{c}{$\circ$}               &
\\
 & & & & & & & & & & & &  \\
\hline
 & & & & & & & & & & & & \\
NGC 628& Sc(s)I & 01 34 0.7& 15 31 55& 656 &10& 25 & 20& 65& 10.7 & 
25 & 6.5 &loose group, comp. at 140 kpc   \\
NGC 3938& Sc(s)I& 11 50 13.6& 44 24 07& 808&10& 11  &4.9& 22 & 5.4 & 
20 & 11.5 & Ursa Mayor Cl., no comp.$<$ 100kpc \\
 & & & & & & & & & & & & \\
\hline
\end{tabular}
\\
\end{flushleft}
\end{table*}

\section{Relevant galaxy properties}

\subsection{NGC 628}

NGC 628 (M74) is a large (Holmberg radius R$_{Ho}$ = 6\amin)
bright face-on galaxy, with a remarkable \hI\ extension, as large as 
3.3 R$_{Ho}$ (Briggs et al 1980). 
Sandage \& Tamman (1975) propose a distance of 19.6 Mpc from the size of
its \hII\ regions, but most authors adopt a distance of 10 Mpc, 
based on a Hubble constant of 75 \kmsmpc. At this distance, 1\amin\,
is about 3 kpc, and our beams are 1.1 kpc and 580 pc in CO(1-0) and
CO(2-1) lines respectively.

Disk morphology and star formation activity were derived by
Natali et al (1992): they fit the I-band light distribution 
by an exponential disk of scale length 4 kpc, and a bulge of 1.5 kpc extent,
which we will use below to interprete the rotation curve. NGC 628 has 
a very modest star formation rate of 0.75 star/yr, which is fortunate, since 
violent stellar activity agitates the interstellar medium, through the
formation of bubbles, jets, stellar winds, and the intrinsic
or un-perturbed z-velocity dispersion
could not be naturally measured. 

 The stellar velocity dispersion has been derived by van der Kruit \& Freeman
(1984): it is 60$\pm$20 \kms at about one luminosity scale-length, and
its evolution with radius is compatible with an exponential decrease,
with a radial scalelength twice that of the density distribution,
as expected if the stellar disk has a constant scale-height with radius.
This result has been derived for several galaxies by Bottema (1993).
 The \hI\ distribution has been observed at Westerbork with 14\asec x 48\asec\,
beam by Shostak \& van der Kruit (1984)
who derived an almost constant z-velocity dispersion (9-10 \kms in the 
center, to 7-8 \kms in the outer parts). Kamphuis \& Briggs (1992)
investigate in more details the \hI\ outer disk with the VLA 
(beam 53\asec x 43\asec).  They found that the inner disk of NGC 628 is
relatively unperturbed, with an inclination of 6.5$^\circ$ and a 
PA of 25$^\circ$, while at a radius of about 6\amin\, (18 kpc), the plane
begins to be tilted, warped and perturbed by high velocity \hI\ clouds .
These clouds could 
be accreting onto the outer parts, and fueling the warp.

As for the molecular component, 
a cross has been observed in NGC 628 by Adler \& Liszt (1989)
with the Kitt Peak 12m telescope (beam 1\amin=3 kpc), and Wakker \& 
Adler (1995) presented BIMA interferometric observations, with 
a resolution of 7.1\asec x 11.6\asec\, (340 x 560 pc). They showed that there
might be a CO emission hole in the center, of size $\approx$ 10\asec, but
this could be only a relative deficiency, since only 45\% of the single-dish
flux is recovered in the interferometric observations.
 The total masses derived are M(H$_2$)=2 10$^9$ M$_\odot$, M(HI)=
1.2 10$^{10}$ M$_\odot$, and from the flat rotation curve V$_{rot}$=200 \kms
an indicative total mass inside two Holmberg radii (12\amin=
36 kpc) M$_{tot}$ =  3.3 10$^{11}$ M$_\odot$, assuming a spherical
mass distribution.

\begin{figure}
\psfig{figure=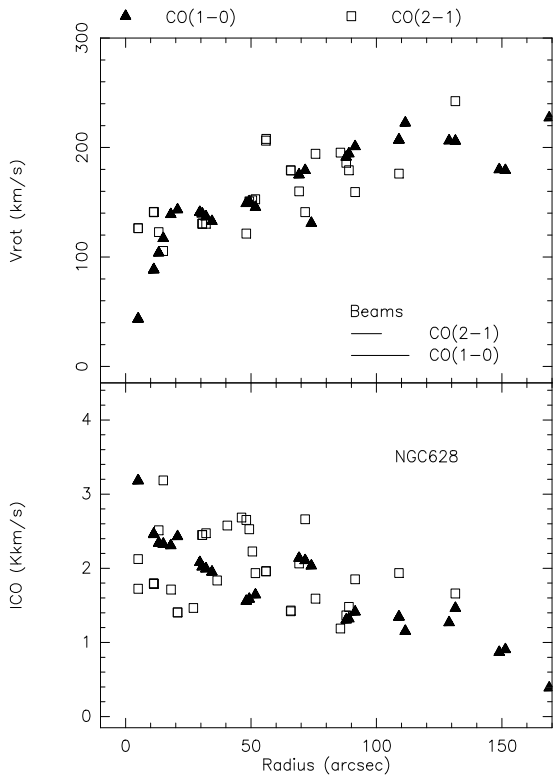,height=12.cm,width=9.cm}
\caption[]{{\bf top} Rotation curve derived from the CO(1-0) (filled triangles)
and CO(2-1) (open squares) observed points in NGC 628. The adopted inclination
is 6.5$^\circ$.
{\bf bottom} Radial distribution of integrated CO emission in NGC 628. }
\label{rad628}
\end{figure}

\begin{figure}
\psfig{figure=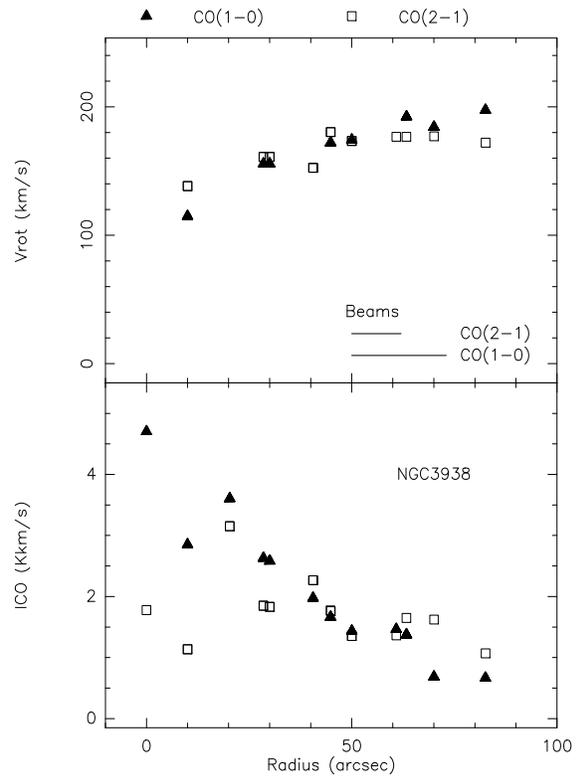,height=12.cm,width=9.cm}
\caption[]{{\bf top} Rotation curve derived from the CO(1-0) (filled triangles)
and CO(2-1) (open squares) observed points in NGC 3938. The adopted inclination
is 11.5$^\circ$.
{\bf bottom} Radial distribution of integrated CO emission in NGC 3938. }
\label{rad3938}
\end{figure}

\subsection{NGC 3938}

NGC 3938 is also a nearly face-on galaxy, at about the same distance
as NGC 628. Its distance derived from its corrected radial velocity
(850 \kms) and a Hubble constant of 75 \kmsmpc is 11.3 Mpc, but Sandage 
\& Tamman (1974) derive a distance of 19.5 Mpc from the \hII\ 
regions. We will adopt here a distance of 10 Mpc, to better compare with 
the literature, where this value is more frequently used.

Its global star formation rate is comparable to that of NGC 628, its ratio
between far-infrared and blue luminosity is L$_{FIR}$/L$_B$=0.15
(while L$_{FIR}$/L$_B$=0.19 for NGC 628).
The stellar velocity dispersion was measured by Bottema (1988, 1993)
to be about 20 \kms at one scale-length: its radial variation is 
also exponential, compatible with a constant stellar scale height.

An \hI\ map was obtained at Westerbork with 24\asec x 36\asec\,  beam
(1.1 x 1.7 kpc) by van der Kruit \& Shostak (1982), and the z-velocity 
dispersion 
was also found almost constant with radius at a value of 10 \kms. They
found no evidence of a systematic pattern of z-motions in the \hI\ layer,
in excess of 5 \kms. CO emission has been reported by Young et al (1995)
towards 4 points, with a beam of 45\asec\, (2.1 kpc), and the central
line-width was 70 \kms. 
 The total masses derived are M(H$_2$)=1.6 10$^9$ M$_\odot$, M(HI)=
1.6 10$^{9}$ M$_\odot$, and from the flat rotation curve V$_{rot}$=180 \kms
an indicative total mass inside two Holmberg radii (7\amin=
21 kpc) M$_{tot}$ =  1.5 10$^{11}$ M$_\odot$, assuming a spherical
mass distribution.

There are some uncertainties about the inclination, as it is usual
for nearly face-on galaxies. Danver (1942) adopted an inclination of 
i=9.5$^\circ$, and van der Kruit \& Shostak (1982) after fitting
the \hI\ velocity field and from the galaxy type deduce i=8$^\circ$-11$^\circ$.
This corresponds to a maximum velocity of 200-250\kms. 
Bottema (1993) uses the Tully-Fisher relation established by
Rubin et al (1985) to deduce a maximum rotational velocity
of 150\kms, corresponding to an inclination of 15$^\circ$.
 However, the Tully-Fisher relation has an intrinsic scatter.
 If we compare with NGC 628, NGC 3938 has about half the 
luminosity, and half the radial scale-length (the exponential
scales of the disk are h$_d$=4 and 1.75kpc, and D$_{25}$ see Table 1). 
If we choose comparable M/L ratios, given they have the same type,
and about the same star formation rate, we expect comparable
maximum rotational velocities. We then adopt a compromise
of V$_{max}\approx$ 180\kms, and an inclination of i=11.5$^\circ$.
 We note that this will not change our conclusions about
the vertical dispersions, except that the critical 
dispersions for stability $\sigma_c$ scale as V$_{max}^{-1}$.

\section{Observations}

Most of the observations were made in August 1994 with the
IRAM 30--m telescope, equipped with single side--band tuned SIS receivers
for both the CO(1-0) and the CO(2-1) lines. The observations were done in
good weather conditions, but the relative humidity was typical of summer
time, which affected essentially the high frequency observations 
(the CO(2-1) line). The typical system temperatures, measured in
the  Rayleigh--Jeans main beam brigthness temperature (T$_A^*$) scale,
of 350\,K and 1000\,K for the 115 and 230 GHz lines, respectively. 

We used two 512 channel filterbanks, with a channel separation of 2.6\,\kms
and 1.3\,\kms for the CO(1-0) and CO(2-1) line respectively.
Each channel width could be slightly broader (by 10 or 20\%) than
the spacing of 1 MHz, degrading slightly the spectral resolution, but
this is not critical, with respect to the half-power line width of at
least 14\kms that we observed.
The half power beam sizes at 115 and 230 GHz are 23\asec\ and
12\asec. The observations were done using a nutating
secondary, with a beamthrow of $\pm$4\amin\ in azimuth. Pointing checks
were done at least every two hours on nearby continuum sources and planets,
with rms fluctuations of less than 3\asec. The calibration of the receivers
was checked by observing Galactic sources (Orion\,A and SgrB2).

The intensities are given here in T$_A^*$,
the chopper wheel calibrated antenna temperature. To convert into main
beam temperatures T$_{mb} =\,{T_{A}^{*} \over \eta_{mb}}\,$
where $\eta_{mb}$ is the main--beam efficiency.
For the IRAM telescope $\eta_{mb}$ is
0.60 and 0.45 for the CO(1-0) and CO(2-1) lines.

For NGC 628, we also combined a series of 
of CO(1-0) and CO(2-1) spectra obtained
during the IRAM galaxy survey by Braine et al (1993). 
These were two radial cuts along the RA and DEC axis. In August 1994, we
observed two radial cuts aligned along the major and minor axis
of the galaxy, with a position angle 25$^\circ$ with respect 
to the previous cuts. We integrated about 30 min per point, and reached
a noise level of about 15 mK and 30 mK for the CO(1-0) and CO(2-1) lines,
at 2.6 \kms resolution. We have smoothed the CO(2-1) data to 2.6 \kms
resolution to increase the signal-to-noise; according to the line-width
observed (at least 10 \kms), this has never a significant broadening
effect (less than 5\%).

Some $^{13}$CO spectra were also taken in November 1994 towards selected
points in NGC 628, to test the effect of cloud overlap, and "macroscopic"
optical depth of the $^{12}$CO line on the derived line-width and
velocity dispersion. The typical system temperatures were then
240 and 350 K at the $^{13}$CO(1-0) and (2-1) (110 GHz and 220 GHz)
respectively. 

\begin{figure}
\psfig{figure=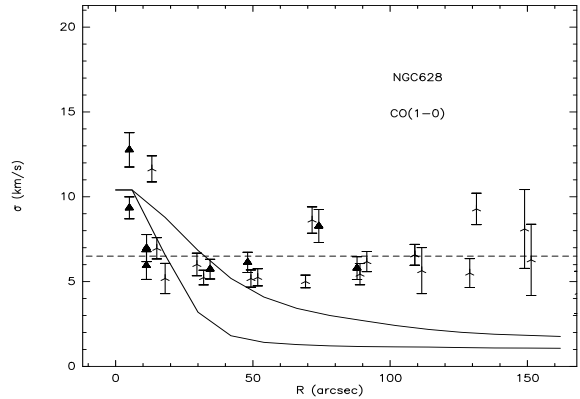,height=6.cm,width=9.cm,angle=-90}
\caption[]{Radial distribution of CO(1-0) velocity dispersions
obtained through gaussian fits of the NGC 628 profiles. The points on
the minor axis are marked by filled triangles. The full lines are the
$\sigma_v$ expected from an axisymmetric velocity models, where the width
comes only from the beam-smearing of the rotational velocity gradients
projected on the sky plane. The top line is for the minor axis, the
bottom line for the major axis. The dashed horizontal line is the
average value adopted for the $^{12}$CO dispersion}
\label{dv_628}
\end{figure}

\begin{figure}
\psfig{figure=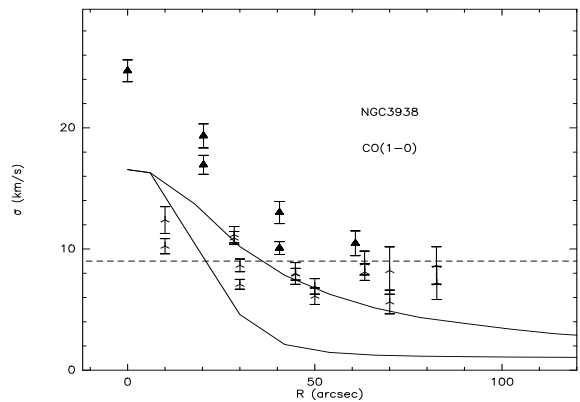,height=6.cm,width=9.cm,angle=-90}
\caption[]{Radial distribution of CO(1-0) velocity dispersions
obtained through gaussian fits of the NGC 3938 profiles. 
Markers and lines as in previous figure.}
\label{dv_3938}
\end{figure}

\section{Results}

\subsection{Spectra and radial distributions}

An overview of the CO(1-0) spectra of the observed galaxies is shown in 
Figures \ref{map628} and \ref{map3938}. We derived the area, central velocity
and velocity width of each profile through gaussian fits. Positions where
the signal-to-noise ratio was not sufficient (below 3) were discarded.
 The radial distributions of integrated CO emission is displayed
in figures \ref{rad628} and \ref{rad3938}, together with the derived
rotation curves. The adopted positions angles and inclinations are
those found for the inner disk in \hI\ by Kamphuis \& Briggs (1992) for
NGC 628, and and a compromise between the values from Bottema (1993) and 
van der Kruit \& Shostak (1982) for NGC 3938. We minimised
the dispersion by fitting the central CO velocity, which is indicated
in Table 1. 

\subsection{Vertical velocity dispersion}

The velocity dispersions, derived directly from
the gaussian fits are displayed as a function of radius in 
figures \ref{dv_628} and \ref{dv_3938} for NGC 628 and NGC 3938 
respectively. The striking feature is the almost constancy of
the velocity dispersion as a function of radius. The dispersion is slightly
increasing towards the center, but this can be entirely accounted for
by the rotational velocity gradients projected on the sky plane.
 We estimated these gradients for each position by modeling 
the radial distribution and velocity field of the molecular gas, assuming
axisymmetry. The radial distribution was taken from the present
observations: we fitted an exponential surface density model for the
CO integrated emission, with an exponential scale-length of 6 kpc
and 3 kpc respectively for NGC 628 and NGC 3938. This corresponds
also to the radial distribution found by Adler and Liszt (1989) for NGC 628.
 We know that there might be a 10\asec\, ($\approx$ 500 pc) hole in the 
NGC 628 center (Wakker \& Adler 1995), corresponding to a hint of depletion
in our CO(2-1) distribution; we have tested this in the model, and
the effect on the expected beam-smoothed line-width was negligible.
  We also entered the values for the rotation curve obtained both from
the inner \hI\ disks (Shostak \& van der Kruit 1984, van der Kruit
\& Shostak 1982), and the present CO rotational velocities. Both 
are compatible (cf fig. \ref{rad628} and \ref{rad3938}). We distributed
randomly 4 10$^4$ test particules according to these radial distributions
and kinematics, in a plane of thickness 100pc. No velocity 
dispersion of any sort was added, so that the spectrum of each
particule is a delta function. After projection, and beam-smearing
with a gaussian beam of 23\asec\, and 12\asec\, to reproduce the CO(1-0)
and CO(2-1) respectively, the expected map of the velocity widths
coming from in-plane velocity gradients was derived. The corresponding
cuts along the minor and major axis are plotted in figures 
\ref{dv_628} and \ref{dv_3938}.

\begin{figure}
\psfig{figure=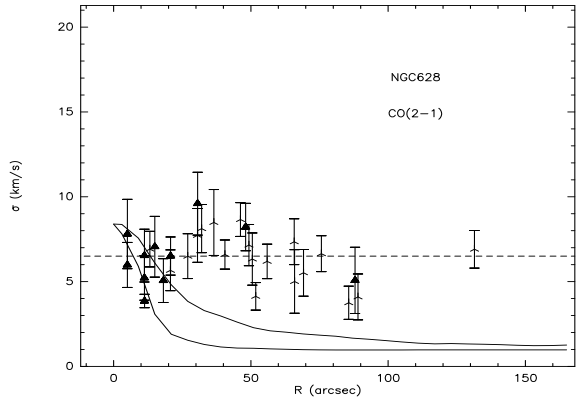,height=6.cm,width=9.cm,angle=-90}
\caption[]{Same as fig \ref{dv_628} but for the CO(2-1) line.}
\label{dv21_628}
\end{figure}

\begin{figure}
\psfig{figure=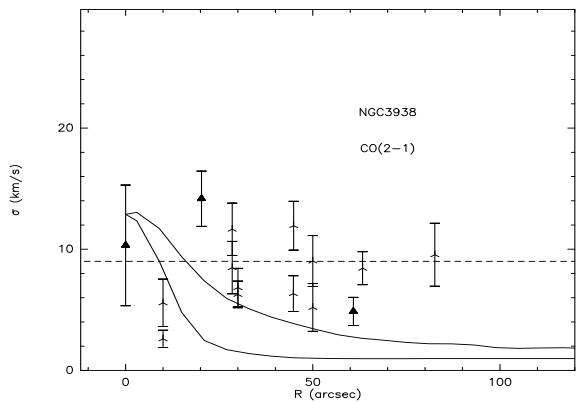,height=6.cm,width=9.cm,angle=-90}
\caption[]{Same as fig \ref{dv_3938} but for the CO(2-1) line.}
\label{dv21_3938}
\end{figure}

We see that in the center, the velocity width can sometimes be entirely
due to the rotational gradients. This is not true for the outer parts, where 
the measured $\sigma$ must represent the vertical velocity dispersion.
 These large effects of the rotational gradients explain the much larger
line-widths found with a 45\asec\, beam by Young et al (1995); they also
explain the spiral shape of the velocity residuals found in the 
\hI\ kinematics by Foster \& Nelson (1985), with comparable spatial 
resolution.

  When trying to deconvolve the measued $\sigma$ at the center, we obtain a 
rather flat profile of vertical velocity dispersion with radius. 
There are however some uncertainties: first, sometimes the expected gradient
is larger than the observed one; this could be explained if the CO-emitting
clouds are not spread all over the beam, and do not share the whole 
expected rotational gradients (for instance, if they are confined into
arms). Also, the axisymmetric model might under-estimate the expected
rotational gradients, since no streaming motions have been taken into
account. We dont estimate it worth to refine the model, 
given the many sources of uncertainty.
  
The beam-smearing is less severe in the CO(2-1) line. The corresponding
curves are plotted in fig \ref{dv21_628} and \ref{dv21_3938}.
Unfortunately, the signal-to-noise ratio is lower for this line,
due to both a higher system temperature, and a CO(2-1)/CO(1-0) emission
ratio slightly lower than 1. The constancy of the line-width as a
function of radius is however confirmed.

\begin{figure}
\psfig{figure=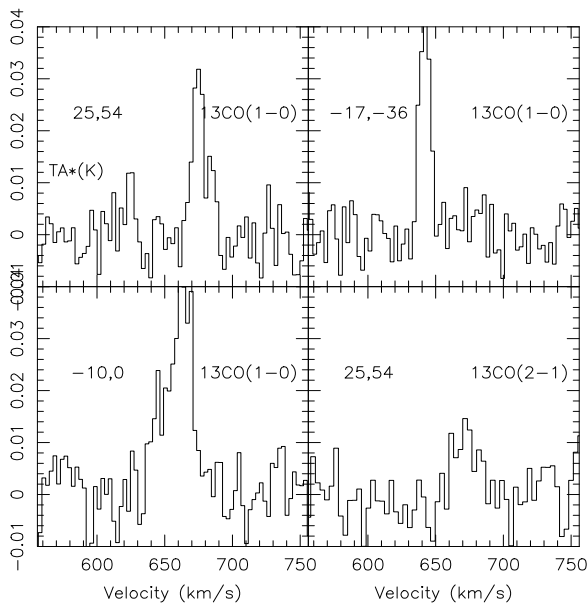,height=9.cm,width=9.cm}
\caption[]{ Some  $^{13}$CO spectra taken towards NGC 628. The offsets
are indicated at the top left, in arcsec. }
\label{co13}
\end{figure}

The constant value of the vertical dispersion 
$\sigma_v$ = (FWHM/2.35) is 6.5 \kms and 9 \kms
for NGC 628 and NGC 3938 respectively. We can question the fact
that this is not the true molecular gas dispersion, if there is a 
saturation effect in the $^{12}$CO line, as shown by Garcia-Burillo
et al (1993). The less saturated  $^{13}$CO line profiles were found
systematically narrower in the galaxy M51. To investigate this point,
we observed a few  $^{13}$CO spectra, as shown in fig \ref{co13}.
  We found indeed slightly narrower $^{13}$CO line profiles, 
with a $^{12}$CO/$^{13}$CO width ratio of 1.1 in average, in the (1-0)
and the (2-1) lines as well. Only the (-17\asec,-16\asec) offset has a
significantly higher ratio width of 1.35 (FWHM = 12.7 $\pm$0.9 for
the  $^{12}$CO line and FWHM=9.1 $\pm$ 0.7 for the  $^{13}$CO line).
  We can therefore conclude that the saturation effect is no more than 
10\% in order of magnitude in average over the galaxy, and we estimate
the true velocity dispersion of the molecular gas perpendicular to the 
plane to be 6 \kms and 8.5 \kms for NGC 628 and NGC 3938.

\subsection{Comparison of the two galaxies}

\begin{figure}
\psfig{figure=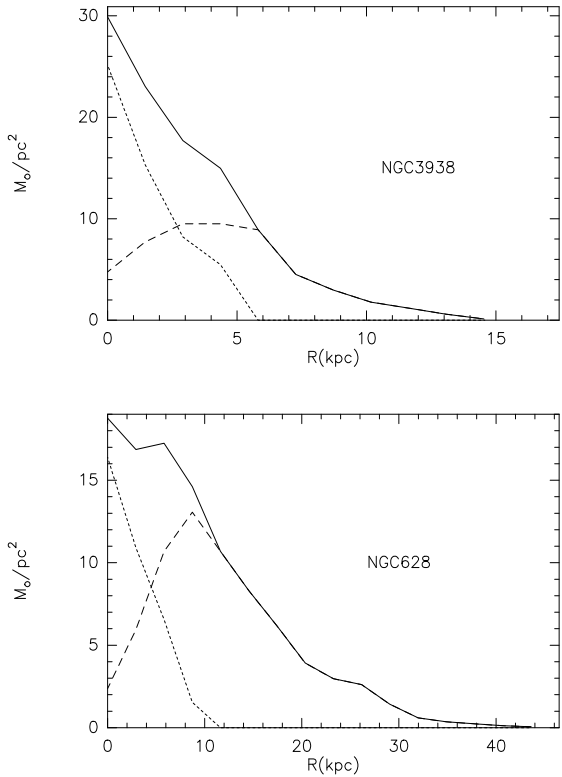,angle=0,height=12.cm,width=9.cm}
\caption[]{ Radial distributions of gas surface densities in NGC 628 (left)
and NGC 3938 (right): \hI\ (dash) and H$_2$ from CO (dots) are combined to
estimate the total (full line) surface density. Helium is taken into
account}
\label{sigma}
\end{figure}

It is interesting to compare the two galaxies NGC 628 and NGC 3938,
since they are of the same type, and however the stellar surface density
is about twice higher in NGC 3938, a property independent of the distance
adopted (both the total luminosity and the characteristic radius are
twice lower in NGC 3938). We can also note that the gas surface
density is also twice higher in NGC 3938 (figure \ref{sigma}).
How can we then explain that the vertical gas dispersion is higher in
NGC 3938, while the stellar vertical dispersion is twice lower? 
First let us note that the maximum rotational velocities
are comparable in the two galaxies, and that is expected if they have
similar M/L ratios: indeed the square of the velocity scales as M/R,
which is similar for both objects.
Now the critical velocity dispersions for stability in the plane
scale as $\sigma_c \propto \mu/\kappa \propto \mu R/V$ and is also
the same for both galaxies. We therefore expect the same planar
dispersions, if they are regulated by gravitational instabilities.
Since we observe a stellar vertical dispersion lower in NGC 3938,
this means that the anisotropy is much higher in this galaxy.

For the self-gravitating stellar disk, we can apply the isothermal
equilibrium, and find that the stellar scale-height $h_*$ scales
as $\sigma_*^2/\mu_*$ and should be 8 times smaller in NGC 3938.
The stellar density in the plane $\rho_0$ should then be 16 times
higher, and so should be the restoring force for the gas $K_z$. We 
can then deduce that the scale height of the gas is also smaller,
but only by a factor 4 or less. In fact, there remains a free parameter, which 
is the vertical/planar anisotropy, which must result from
the history of the galaxy formation and evolution
(companion interactions, mergers, gas infall etc..)
which should explain the differences between the two galaxies.

\section{Summary and Discussion}

 One of the most important result is that the vertical velocity
dispersion of the molecular gas in NGC 628 and NGC 3938
is constant as a function of radius. Morevover
the value of this dispersion is $\sigma_v \approx$ 6-8 \kms, very
comparable to that of the \hI\ component.

  This universality of the dispersion already tells us that it does
not correspond to a thermal width. In that case, the dispersion
is only a function of gas temperature, and this should vary with 
the galactocentric distance, since the cooling and heating processes
depend strongly on the star formation efficiency. The temperature
of dust derived from infra-red emission for instance, is a function
of galactocentric radius. Also, the dispersion of the \hI\ should then
be very different from that of the colder molecular gas.
  It is clear at least for the cold molecular component that the line widths
correspond to large-scale macroscopic turbulent motions between
a large number of clumps of internal dispersion possibly 
much lower than 1 \kms.

 The fact that the \hI\ and CO emissions reveal comparable z-dispersions
may appear surprising. It is well known from essentially
our own Galaxy, that the \hI\ plane is broader than the molecular gas plane,
and this is attributed to a larger z-velocity dispersion
(e.g. Burton 1992). In external galaxies, the evidence is indirect,
because of the lack of spatial resolution. In M31, the comparison
between \hI\ and CO velocity dispersions led to the conclusion that
the \hI\ height was 3 or 4 times higher than the molecular one at
R=18 kpc (Boulanger et al 1981). Is really the thin molecular component
an independant layer embedded in a thicker \hI\ layer?
We discuss this further below.

\subsection{Critical dispersion for gravitational stability}

It is interesting to compare the observed vertical
velocity dispersions to the critical dispersion required in the plane 
by stability criteria; as a function of gas surface 
density $\mu_g$ and epicyclic frequency $\kappa$, the critical
dispersion can be expressed by (Toomre 1964):
$$
\sigma_{cg} = 3.36 G \mu_g/\kappa
$$
To obtain the total gas surface density $\mu_g$, we have combined the available
\hI\ data (Shostak \& van der Kruit 1984 for NGC 628; van der Kruit \& Shostak 
1982, for NGC 3938) to the present CO data as a tracer of H$_2$ surface density.
 We have multiplied the result by the factor 1.4, to take into account
the helium fraction.
 We can see in figure \ref{sigma} that the apparent central hole detected
in \hI\ is filled out by the molecular gas.
This is not unexpected, since it is believed that above a certain gas
density threshold, the atomic gas becomes molecular.
 This threshold involves essentially the gas column density, the pressure
of the interstellar medium, the radiation field and the metallicity, since the 
main point is to shield molecules from photodestruction (Elmegreen 1993). It 
is indeed observed that the average gas column density is sufficiently 
increasing towards the galaxy centers, to reach the threshold. Once the 
shielding conditions are met, the chemistry time-scale is short enough (of 
the order of 10$^5$ yrs) with respect to the dynamical time-scale, that the
\hI\ to H$_2$ phase transition occurs effectively. This 
transition is obvious in most galaxies (e.g. Sofue et al 1995,
Honma et al 1995); the threshold at solar conditions for metallicity and 
radiation field is around 10$^3$ cm$^{-3}$ and 10$^{21}$ cm$^{-2}$, but
it is difficult to precise it more because of
spatial resolution effects, and because we rely upon CO emission
to trace H$_2$ (the observed thresholds concern
in fact CO excitation and photo-dissociation).
\medskip

We have then built a mass model of the two galaxies in fitting their 
rotation curves, taken into account the constraints on the scale-lengths
and masses of luminous components given by optical observations
(section 2). In these fits, the gas contribution to the rotation curve
have been found negligible. We include in the mass model a spherical bulge
represented by a Plummer (size r$_b$, mass M$_b$), an exponential
disk (scale-length h$_d$, mass M$_d$) and a flattened dark matter halo,
represented by an isothermal, pseudo-ellipsoidal density
(eg Binney \& Tremaine 1987):
\begin{eqnarray*}
\rho_{DM} & = & \frac{\rho_{0,DM}}{(1+ \frac{r^{2}}{r_h^{2}} + 
              \frac{z^{2}}{r_h^{2} q^{2}})} 
\end{eqnarray*}
where $r_h$ is the characteristic scale of the DM halo and $q$ its 
flattening.
All parameters of the fits are displayed in Table 2. 
The fits are far from unique, but their essential use is to get
an analytical curve fitting the observed rotation curve,
in order to get derivatives and characteristic dynamical frequencies.
Given the functional forms adopted for the various components,
we can get easily the total mass, and compare the M/L with
expected values for galaxies of the same types.

\begin{table}
\begin{flushleft}
\caption[]{Mass models derived from rotation curve fits}
\begin{tabular}{|lcc|}
\hline
  & & \\
\multicolumn{1}{|l}{Galaxy}             &
\multicolumn{1}{c}{NGC 628}              &
\multicolumn{1}{c|}{NGC 3938}       \\
& &  \\
\hline
 & & \\
r$_b$(kpc)  &  1.5    &  0.8      \\
M$_b$(10$^{10}$ M$_\odot$)  &  1.6    &  0.8      \\
h$_d$(kpc)  &  3.9    &  1.75      \\
M$_d$(10$^{10}$ M$_\odot$)  &  6.9    &  2.8      \\
r$_h$(kpc)  &  14.    &  7.      \\
M$_h$(10$^{10}$ M$_\odot$)$^*$  &  2.7    &  1.3      \\
$q$  &  0.2    &  0.2     \\
M(stars)/L$_B$ &  3.4  & 3.3 \\
M$_{tot}(<R_{25})$/L$_B$ &  4.5  & 4.5 \\
 & & \\
\hline
\end{tabular}
\\
\vskip 4truemm  
$^*$ Mass inside R$_{25}$=15.5 kpc for NGC 628 and 7.85 kpc for NGC 3938\\
\end{flushleft}
\end{table}

\begin{figure}
\psfig{figure=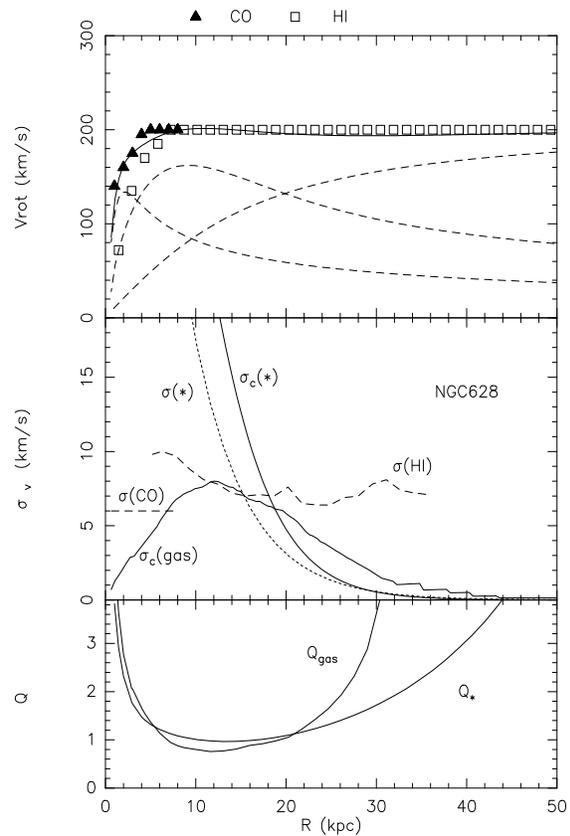,height=12.cm,width=9.cm}
\caption[]{ Rotation curve fit for NGC 628:
{\it top}: total fitted rotation curve (full line), with contributions
of bulge, exponential disk and dark matter halo (dashed lines) compared
with CO and \hI\ data;
{\it middle}: Derived critical velocity dispersions required for
axisymmetric stability for stars and gas (full lines). The \hI\ and CO observed 
vertical velocity dispersions are also shown for comparison (noted $\sigma$(HI) 
full line and $\sigma$(CO), horizontal dashed line). The \hI\ dispersions
data have been taken from the compilation in Kamphuis thesis (1992).
A fit to the observed 
z-stellar velocity dispersion is also shown ($\sigma$(*), dotted line);
{\it bottom}: Corresponding Toomre $Q$ parameters, assuming
$\sigma_z/\sigma_r$ =0.6 for the stars (the CO dispersion has
been taken for the gas)}
\label{vrot628}
\end{figure}

\begin{figure}
\psfig{figure=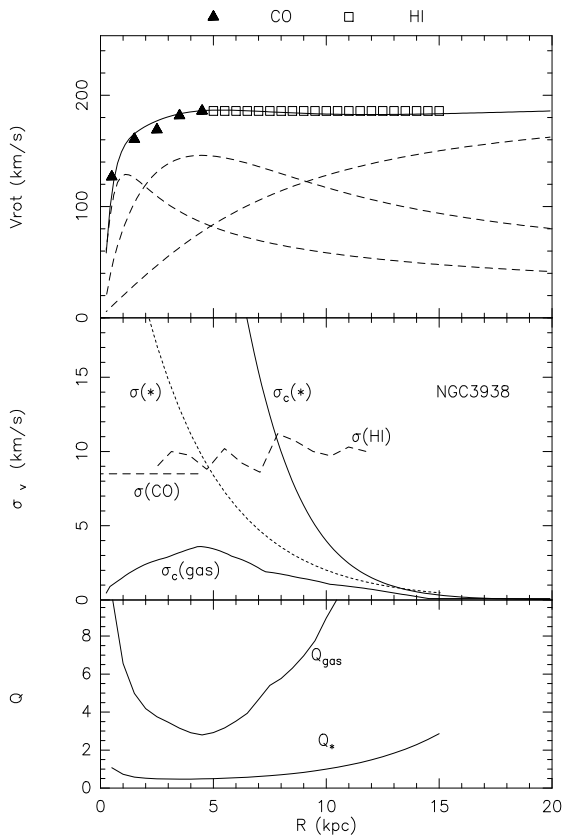,height=12.cm,width=9.cm}
\caption[]{ Same as previous figure, for NGC 3938}
\label{vrot3938}
\end{figure}

From these mass models, we have derived the epicyclic frequency as a function
of radius (this does not depend on the precise model used,
as long as the rotation curve is fitted), and the critical velocity
dispersion required for axisymmetric stability,
for the stellar and gaseous components (figures \ref{vrot628} and
\ref{vrot3938}). The comparison with the observed vertical velocity
dispersions for \hI\ and CO is clear: the observed values are most of
the time larger, in particular for NGC 3938. This means that, if the gas
velocity dispersion can be considered isotropic, the Toomre stability
parameter in the galaxy plane
is always $Q_g \ga 1$, and most of the time  $Q_g >$ 2-3, for NGC 3938.
For NGC 628, $Q_g$ is near 1 between 3 and 20kpc, and the threshold
for star formation, $Q_g=1.4$ according to Kennicutt (1989) is
reached at 23 kpc. This is far in the outer parts of the galaxy,
since R$_{25}$ = 15.5 kpc. 

If the vertical dispersion is lower than in the plane, as could be
the case (e.g. Olling 1995), than $Q_g$ is even larger.
 The gas appears then to be quite stable, unless the coupling
gas-stars has a very large effect.

Figures \ref{vrot628} and \ref{vrot3938} also plot the critical
velocity dispersion for the stellar component, together with a fit
to the observed stellar velocity dispersions, from van der Kruit \&
Freeman (1984) for NGC 628 and from Bottema (1988, 1993) for NGC 3938.
From a sample of 12 galaxies where such data are available,
Bottema (1993) concludes that the stellar velocity dispersion is
declining exponentially as $e^{-r/2h}$, as expected for an exponential
disk of scale-length $h$ and constant thickness, as found by 
van der Kruit \& Searle (1981). Since mostly the vertical stellar
dispersion $\sigma_z$ is measured, it is assumed that there is a constant ratio
between the radial dispersion $\sigma_r$, comparable to that observed
in the solar neighbourhood $\sigma_z/\sigma_r$ = 0.6. This is already
well above the minimum ratio required for vertical stability, i.e.
$\sigma_z/\sigma_r$ = 0.3 (Araki 1985, Merritt \& Sellwood 1994).
Within these assumptions, it can be derived that the Toomre parameter
for the stars $Q_*$ is about constant with radius, within the 
optical disk; it depends of course
on the mass-to-light ratio adopted for the luminous component, and
is in the range  $Q_* \approx$ 1 for M(stars)/L$_B$ = 3. 
 Figures \ref{vrot628} and \ref{vrot3938} confirm the result of almost constant
$Q_*$, but with low values, especially for NGC 3938. 
This could be explained, if the vertical dispersion is indeed much lower
than the radial one. The minimum value for the ratio
$\sigma_z/\sigma_r$ is 0.3 (for stability reasons), so that the derived 
$Q_*$ values displayed in figures \ref{vrot628} and \ref{vrot3938} could be 
multiplied by $\approx$ 2. The idea of
stellar velocity dispersion regulated by gravitational instabilities
appears therefore supported by the data, within the uncertainties.
\medskip

The most intriguing result is the large gas vertical dispersion observed
for NGC 3938, and its distribution with radius. The large corresponding $Q_g$
values, that will mean comfortable stability, are difficult to reconcile with 
the observed large and small-scales gas instabilities: clear spiral arms
are usually observed in the outer \hI\ disks, with small-scale structure as well
(see e.g. van der Hulst \& Sancisi 1988, Richter \& Sancisi 1994). This
is also the case here for NGC 628 showing all signs of gravitational
instabilities in its outer \hI\ disk (Kamphuis \& Briggs 1992), and 
for NGC 3938 (van der Kruit \& Shostak 1982).
A possibility to reduce $Q_g$ is that also the gas dispersion
is anisotropic, this time the vertical one being larger than in the plane.
However we will see, through comparison with
gas dispersion in the plane of the Galaxy (cf next section) 
that the anisotropy of gas dispersion does not appear so large.
Another explanation could be that 
the present rough calculations of the $Q$-parameter concern only
a simplified one-component stability analysis, and could be
significantly modified by multi-components analysis. 
It has been shown (Jog \& Solomon 1984, Romeo 1992, Jog 1992 \& 1996) 
that the coupling between
several components de-stabilises every dynamical component.
 The apparent stability ($Q\approx 2-3$) of the gas component
might therefore not be incompatible with an instability-regulated
velocity dispersion for the gas.

But then, in the vertical direction, the dispersion is much higher
than the minimum required for vertical stability.
 Could this large velocity dispersion be powered by star formation?
This is not likely, at least for the majority of the \hI\ gas
well outside the optical disk, where no stellar activity is observed.
A possible explanation would be to suppose that the \hI\ is tracing a much larger
amount of gas, in the form of molecular clouds, which will then 
be self-gravitating, with $Q_g \la 1$ (Pfenniger et al 1994; 
Pfenniger \& Combes 1994).
With a flat rotation curve, and a gas surface density decreasing as $1/r$,
the critical dispersion would then be constant with radius. 

\subsection{ Similar \hI\ and CO vertical dispersions}

\subsubsection{ Two phases of the same dynamical component }

 Another puzzle is the similarity of the CO and \hI\ vertical velocity 
dispersions. If the gas layers are indeed isothermal in z, we can deduce
that both atomic and molecular layers have also similar heights.
 This means that the atomic and molecular components can be considered as a 
unique dynamical component, which can be observed under two phases, 
according to the local physical conditions (density, excitation temperature, 
etc..). The amplitudes of z-oscillations of the molecular and atomic
gas are the same, only we see the gas as molecular when it is at
heights lower than $\approx$ 50pc. At these heights, 
the molecular fraction is 
$f_{mol} \ga 0.8$ (Imamura \& Sofue 1997), which means that almost
all clouds are molecular, taking into account their atomic envelope.
In fact it is not clear whether we see the CO or H$_2$ formation and 
destruction, since we can rely only on the CO tracer. Also, it is 
possible that the density of clouds at high altitude is not enough to
excite the CO molecule, which means that the limit for observing CO
will not be coinciding with the limit for molecular presence itself.
The latter is strongly suggested by the observed vertical density profiles
of the H$_2$ and \hI\ number density: there is a sharp boundary where
the apparent $f_{mol}$ falls to zero, while we expect a smoother profile
for a unique dynamical gas component.

That the gas can change phase from molecular to atomic and vice-versa
several times in one z-oscillation is not unexpected, 
since the time-scale of molecular formation and destruction
is smaller than the z-oscillation period, of $\approx$ 10$^8$ yrs
at the optical radius: the chemical time-scale
is of the order of 10$^5$ yrs (Leung et al 1984, Langer \& Graedel 1989). 
Morever, as discussed in the previous section ({\it 5.1}), the key factor 
controlling the presence of molecules is photodestruction, which
explains why there is a column density threshold above which the
gas phase turns to molecular (Elmegreen 1993). This threshold could be
reached at some particular height above the plane.
\bigskip

\subsubsection{ Collisions }

Should we expect the existence of several layers of gas at different
tmperatures, and therefore different thicknesses, in galaxy planes?
In the very simple model of a diffuse and homogeneous gas, 
unperturbed by star-formation, we
can compute the mixing time-scale of two layers at different temperatures,
through atomic or molecule collisions: this is of the order of the 
collisional time-scale, $\approx$ 10$^4$ yrs for an average volumic density
of 1 cm$^{-3}$, and a thermal velocity of 0.3 \kms. This is very short
with respect to the  z-oscillation time scale of $\approx$ 10$^8$ yrs, 
and therefore mixing should occur, if differential dissipation or
gravitational heating is not taken into account.

This simple model is of course very far from realistic.
 We know that the interstellar medium, atomic as well as
molecular, is distributed in a hierachical ensemble of clouds,
similar to a fractal. Let us then consider another simple modelisation
of an ideal gas where the particles are in fact the interstellar
clouds, undergoing collisions (cf Oort 1954, Cowie 1980). 
For typical clouds of 1pc size, and 10$^3$ cm$^{-3}$ volumic density,
the collisional time-scale is of the order of 10$^8$ yrs,
comparable with the vertical oscillations time-scale. 
This figure should not be taken too seriously, given the
rough simplifications, but it corresponds to what has
been known for a long time, i.e. the ensemble of clouds cannot
be considered as a fluid in equilibrium, since the collisional
time-scale is comparable to the dynamical time, 
like the spiral-arm crossing time (cf Bash 1979, Kwan 1979, Casoli \&
Combes 1982, Combes \& Gerin 1985).

If the collisions were able to redistribute the kinetic
energy completely, there should be equipartition, i.e.
the velocity dispersion would decrease with the mass $m$ of the
clouds like $\sigma_v \propto m^{-1/2}$. In fact the cloud-cloud 
relative velocities are roughly constant with mass
(between clouds of masses 100 M$_\odot$ and GMCs of 10$^6$ 
M$_\odot$, a ratio of 100 would be expected in velocity dispersions,
which is not observed, Stark 1979). Towards the Galactic anticenter, where 
streaming motions should be minimised, the one-dimensional dispersion
for the low-mass and giant clouds are found to be about 9.1 and 6.6 \kms 
respectively, with near constancy over several orders of magnitude,
and therefore no equipartition of energy (Stark 1984). 
The almost constancy of velocity
dispersions with mass requires to find other mechanisms
responsible for the heating.

\subsubsection{ Gravitational heating }

 If relatively small clouds can
be heated by star-formation, supernovae, etc...(e.g.
Chi\`eze \& Lazareff 1980), the largest clouds could be
heated by gravitational scattering (Jog \& Ostriker 1988,
Gammie et al 1991). In the latter mechanism, encounters between
clouds with impact parameters of the order of their tidal radius
in a differentially rotating disk are equivalent to a gravitational
viscosity that pumps the rotational energy into random cloud
kinetic energy. A 1D velocity dispersion of 5-7\kms is the predicted 
result, independent of mass. This value is still slightly lower than
the observed 1D dispersion of clouds observed in the Milky Way.
Stark \& Brand (1989) find 7.8\kms from a study within 3 kpc of the
sun. But collective effects, gravitational instabilities forming
structures like spiral arms, etc... have not yet been taken into account.
 Given the high degree of structure and apparent permanent
instability of the gas, they must play a major role in the heating,
the source of energy being also the global rotational energy.
Dissipation lowering the gas dispersion continuously maintains
the gas at the limit of instability, closing the feedback loop
of the self-regulation (Lin \& Pringle 1987, Bertin \& Romeo 1988).
 In the external parts of galaxies, where there is no star formation,
 gravitational instabilities are certainly the essential heating
mechanism. This again will tend to an isothermal, or more exactly
isovelocity, ensemble of clouds, since the gravitational
mechanism does not depend on the particle mass. 
The molecular or atomic gas are equivalent in this process,
and should reach the same equilibrium dispersion.

\medskip

\subsubsection { \hI\ and CO velocity dispersions in the Milky Way }

In the Milky Way, although the kinematics of gas is much complicated due to
our embedded perspective, we have also the same puzzle. 
The velocity dispersion has been estimated through several methods,
with intrinsic biases for each method, but essentially the dispersion
has been estimated in the plane.
Only with high-latitude molecular clouds, can we have an idea of the
local vertical velocity dispersion. Magnani et al (1996) have recently made
a compilation of more than 100 of these high-latitude clouds. The velocity
dispersion of the ensemble is 5.8\kms if seven intermediate velocity objects
are excluded, and 9.9 \kms otherwise. This is interestingly close to
the values we find for NGC 628 (6\kms) and NGC 3938 (8.5\kms). Unfortunately
there is always some doubt in the Galaxy that all molecular clouds are
taken into account, due to many selection effects, while the measurement
is much more direct at large scale in external face-on galaxies.
 In fact, it has been noticed by Magnani et al (1996) that there were an
inconsistency between the local measured scale-height of molecular
clouds (about 60pc) and the vertical velocity dispersion. However, they
conclude in terms of a different population for the
local high-latitude clouds (HLC). Indeed, the
total mass of observed HLC is still a small fraction of the molecular
surface density at the solar radius.

The local gaussian scale height of the molecular component has
been derived to be 58pc (at R$_\odot$ = 8.5kpc) through
a detailed data modelling by Malhotra (1994); this is
also compatible with all previous values (Dame et al 1987, Clemens
et al 1988). The local \hI\ scale height is 220pc (Malhotra 1995).
We therefore would have expected a ratio of 3.8 between the dispersions
of the H$_2$ and \hI\ gas, but these are very similar, within the uncertainties,
which come mainly from the clumpiness of the clouds for the H$_2$ 
component. If we believe the more easily determined \hI\ dispersion
of 9\kms (Malhotra 1995), then the H$_2$ dispersion is expected to
be 2.4\kms, clearly outside of the error bars or intrinsic scatter:
the value at the solar radius is estimated at 7.8\kms by Malhotra (1994). 
Of course, all this discussion is hampered by the fact that we discuss
mainly horizontal dispersions in the case of the Milky Way, while
the gas dispersions could well be anisotropic. This is why the present results
on external face-on galaxies are more promising.

\bigskip

 The vertical gas velocity dispersion in spiral galaxies
is an important parameter required to determine the flattening of
the dark matter component, combined with the observation of the gas
layer thickness (cf Olling 1995, Becquaert \& Combes 1997). We have shown here 
that the gas dispersion does not appear very anisotropic, in the sense
that the vertical dispersion is not much smaller that what has been 
derived in the plane of our Galaxy (for instance by the terminal 
velocity method, Burton 1992, Malhotra 1994). Such vertical dispersion data 
should be obtained in much larger samples, to consolidate 
statistically this result.

\vspace{0.25cm}

\acknowledgements
 We are very grateful to the referee, R. Bottema, for his interesting
and helpful comments. We acknowledge the support from the staff
at the Pico Veleta during the course of these observations.

\end{document}